# COMPARISON OF TWO LOW-POWER ELECTRONIC INTERFACES FOR CAPACITIVE MEMS SENSORS

*G. Nagy[1], Z. Szucs[1], S. Hodossy[2], M. Rencz[1], A. Poppe[1]*

[1]Budapest University of Technology and Economics
{nagyg | szucs | rencz | poppe}@eet.bme.hu
[2]ekrator@gmail.com

**ABSTRACT**

The paper discusses the importance and the issues of interfacing capacitive sensors. Two architectures applicable for interfacing capacitive sensors are presented. The first solution was designed to interface a capacitive humidity sensor designed and built for a humidity-dependent monolithic capacitor developed at Budapest University of Technology and Economics. The second case presents the possible read-out solutions for a SOI-MEMS accelerometer. Both of the architectures were built and tested in a discrete implementation to qualify the methods before the integrated realization. The paper presents a detailed comparison of the two methods.

## 1. INTRODUCTION

Micromechanical sensors are more and more widespread in today's electrical sensing applications. The reason for this is the fact that they are small in size and their power consumption is low. A considerable part of these sensors uses capacitive conversion methods for sensing, because this way low current consumption and high resolution are easily achievable. Another particular advantage of capacitive sensors is the ease of integration with the surrounding circuitry. Unlike their piezoresistive and optical counterparts, most of the capacitive sensors do not require any special additional material or processes beyond those of the basic CMOS technology. Capacitive sensors also have a good stability and a near-zero temperature coefficient.

Capacitive sensors can be applied in a very wide range of applications including gas, pressure, acceleration and humidity sensors. They can also detect motion and flow. Our research group at BUTE is concerned with the development, production, testing and the reliability aspects of gas, humidity and acceleration sensors.

The development of the read-out circuitry for this kind of sensors is however a great challenge as generally a very small signal has to be detected in an extremely noisy environment. Furthermore, calibration, temperature compensation, self-test and analog-to-digital conversion are also necessary in a large number of applications. In order to reduce the size and production cost of the circuit the functions above are often integrated in the sensor's interface. By means of a smart design of the surrounding electronics, the requirements for the mechanical parts can be simplified. Several different architectures exist, the choice amongst which depends on the properties of the mechanical sensing element. This paper deals with the design issues and the selection criteria of the electronic interface for a particular sensor element.

The task of the capacitive sensors' interface is to convert any changes in the value of the capacitor into a variation of a variable that is easy to process by electrical means – e.g. voltage, current, frequency or pulse width. Several methods exist, that can be grouped as follows:

1. Probably the simplest way is to charge the capacitor with a constant current for a certain time and measure its voltage [6]. A very precise current source, time and voltage measurement and an environment with very low noise is necessary for this method, thus it is of little practical use when the capacitance to measure is very small and the environment is noisy. A sophisticated and improved version of this method was used with success by Daniela De Venuto and Bruno Riccò [2].
2. The capacitor can be a part of an RC oscillator the frequency of which will change with the capacitance [8, 11]. This method – as will be shown in this paper – is very easily realizable and is more tolerant to noises than the previous one. High accuracy can not be achieved though and it is not beneficial for small capacitance values.
3. The AC impedance of the capacitor can be measured with a sine-wave source and by measuring the current and the voltage of the capacitor. Very accurate results can be achieved, but the resulting circuitry is extremely complex [6].
4. A very common method for interfacing low-capacitance sensors is to use a charge amplifier that converts the ratio of the sensing capacitor and a reference capacitor into voltage [3, 6].
5. The sigma-delta conversion, a well proven technology for ADCs, can be used to interface capacitive sensors [5, 6]. Its greatest advantage is that it directly converts capacitance to a digital value. According to [5], an outstandingly high resolution (2 aF/$\sqrt{Hz}$), high linearity (100 ppm), and high accuracy (4 fF) can be achieved.

6. As the measurands of the capacitive sensors usually change rather slowly, 1/f noise is a serious concern for the interfaces. The chopper stabilization method [4, 9, 10, 12, 13] offers a solution for this problem. This method will be discussed in details in this paper.
7. Correlated double sampling [4, 9] is another method to decrease the 1/f noise. The sensor interface takes samples from the "noise" (i.e. the amplifier's output, when no input is present) and the noisy signal. This way a very effective noise-cancellation is attainable.

## 2. TWO CASE STUDIES

Two different types of sensing capacitors (a humidity sensor capacitor and a capacitive accelerometer) were chosen for investigation. Their sensing capacitances differ in orders of magnitude, thus different read-out architectures are needed. The structure of the capacitors and the schematic build-up of the interface circuitry are presented below.

### 2.1. The structure and the interface of the humidity sensor

The first sensor in question was based on a humidity sensing capacitor integrated onto a single-crystal silicon chip manufactured at BUTE by wet anodic oxidizing process. Its capacitance is proportional to relative humidity (RH).

The capacitor has a dielectric layer made of porous aluminum-oxide (ceramics). The permittivity of a porous material changes with the humidity thus the capacitance of such a capacitor will be humidity dependent. The bottom electrode is the bulk silicon itself. The dielectric is a layer of porous ceramics and the upper electrode is a very thin layer of porous palladium.

This kind of capacitive sensors can be realized by additional post-CMOS steps after the preproduction of the read-out electronics. The capacitance of such a device is in the range of 180 – 500 pF depending on the RH value.

For the discrete realization of the smart sensor we have developed, designed and manufactured a sensing circuitry based on a *capacitance-to-frequency converter*. A simplified schematic of the circuitry can be seen in Figure 1.

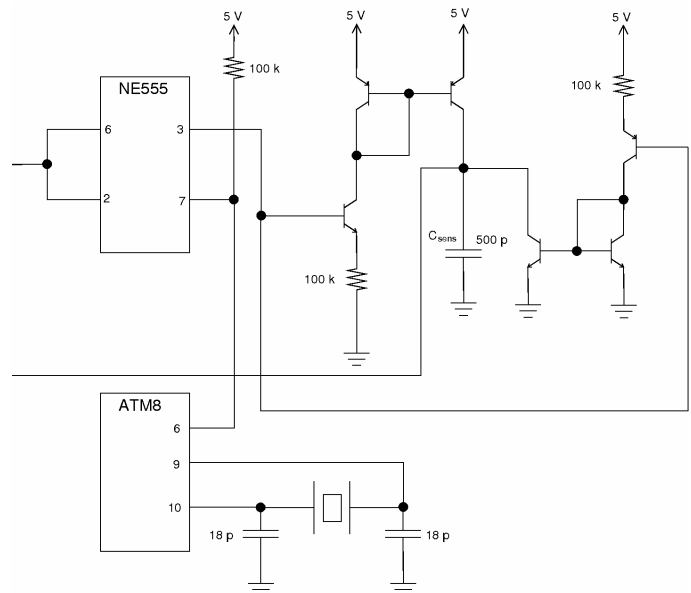

Figure 1 – Simplified schematic of the humidity sensor's read-out circuitry

A 555 timer IC, complemented with a symmetrical charge and discharge current source, is used for the capacitance-to-frequency conversion and the cycle time is measured and digitalized by a microcontroller (ATM8).

The capacitance is charged and discharged with a constant current of 40 µA. The capacitance's voltage is compared to the one-third and the two-thirds of the power supply voltage. Thus the cycle time can be calculated as follows:

$$T = \frac{2}{3}\frac{V_S}{I}(C_0 + C_S) \qquad (1)$$

where $V_S$ is the supply voltage, I is the charging current and $C_S$ is the capacitance of the sensor element. As the equation between T and $C_S$ is linear, an additional capacitor ($C_0$) can be inserted parallel to the sensor to adjust the frequency to a convenient range. The advantage of the lower operating frequency is that the heat dissipation of the sensing capacitor can be considerably reduced.

As this measurement is based on the count of the cycles, it is relatively slow, but has a lower sensitivity to noises. In a discrete realization the parasitic capacitances of the elements and the printed circuit board itself affect the measurement and decrease the accuracy. An integrated implementation of this sensor yields better performance in terms of accuracy, noise distortion and power consumption.

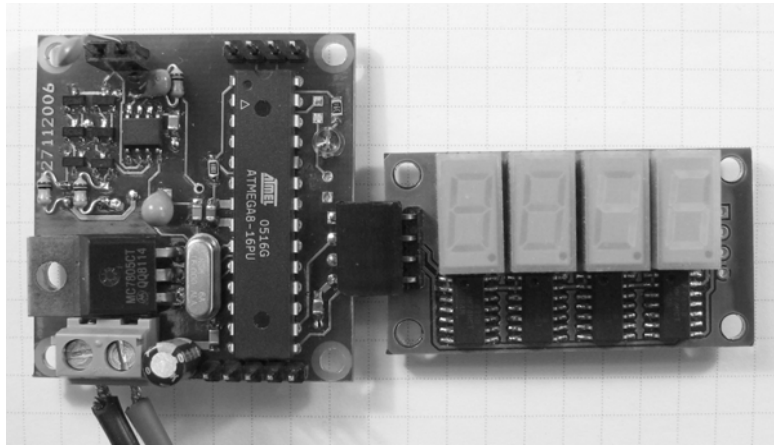

Figure 2 – The humidity sensor

Figure 3. shows the sensitivity of the read-out circuit for different sensing capacitance values in the range of 18 pF to 1 nF. The characteristic was found to be linear as expected.

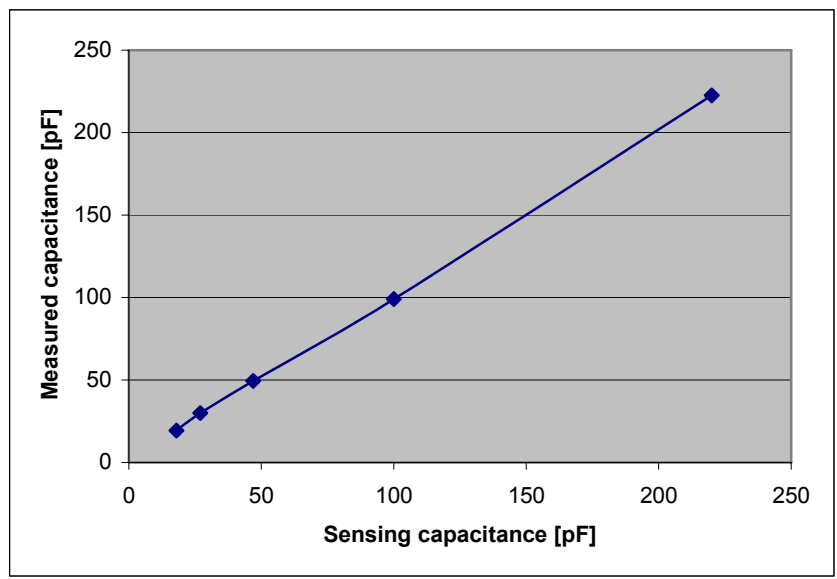

Figure 3 – The characteristics of the read-out circuitry

### 2.2. Interfacing an accelerometer

Accelerometers are generally realized as MEMS combdrive structures forming differential capacitive elements, as shown in Figure 4. The arrangement is built up of one movable and two static plates creating two differential capacitors, a capacitive half-bridge. The displacement of the moving electrode results in a change in the capacitance (ΔC).

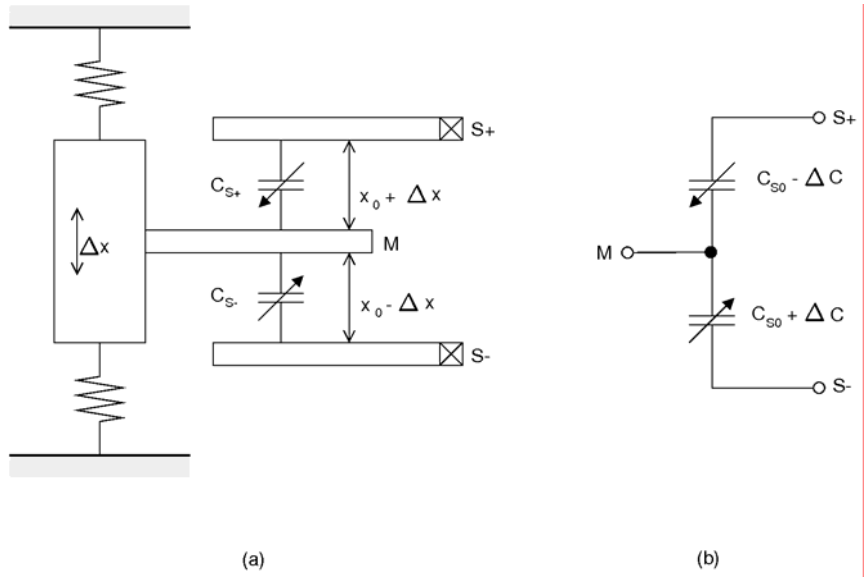

Figure 4 – The principle of differential capacitive sensing

The capacitance-change to be measured is usually a very small value (less than 1 pF) compared to that of the humidity sensor. Consequently this kind of sensor produces a very small output signal that may need amplification of several orders of magnitude. As the frequency of the signal is rather low, Flicker-noise (or 1/f noise) poses a significant problem. To overcome this difficulty, an architecture very different from the previously outlined should be chosen.

In a noisy environment *chopper stabilization* is an effective analog method that gives fairly good results and allows for both discrete and monolithic realization. This method separates the signal from the noise in the frequency domain using amplitude modulation and demodulation.

A sinusoidal voltage of $f_M$ frequency ($v_M$) is applied to the differential MEMS capacitors in the circuit shown in Figure 5. The gain of the amplifier depends on the value of the sensing capacitors, thus the change of their capacitance appears as a change in the amplitude of the amplifier's output signal ($v_y$). This way the signal is transferred to the side-bands of the carrier signal's frequency ($f_M$), while the Flicker-noise is still located at low frequencies. After demodulation using an analog multiplier, the signal and the noise change places in the frequency domain which allows for the removal of the noise from $v_z$ with the help of a low-pass filter ($v_{out}$).

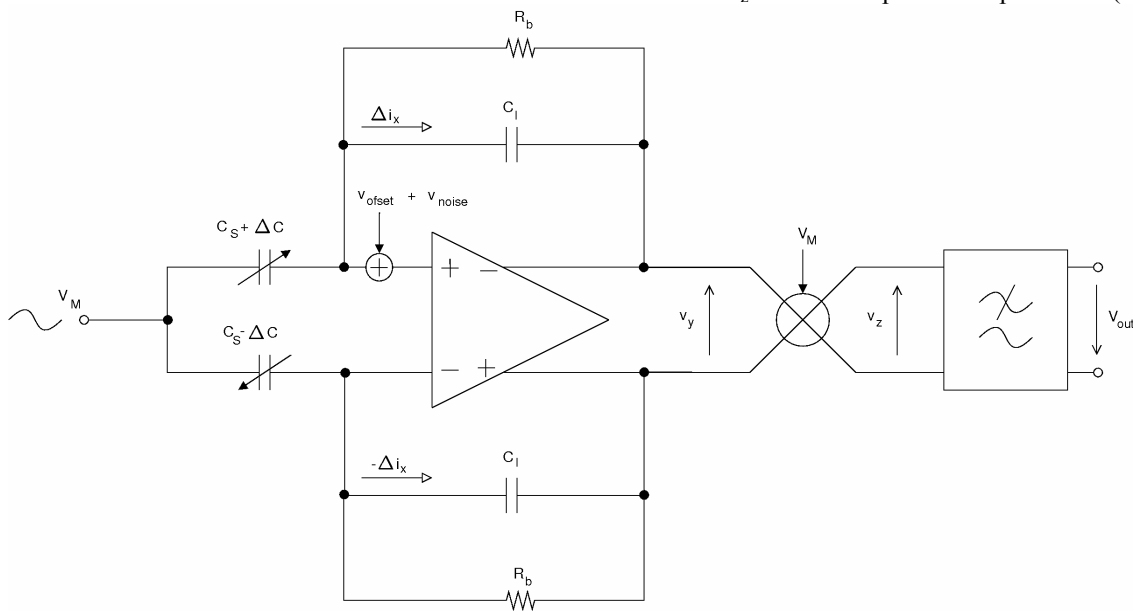

Figure 5 – Chopper stabilization

The chopper stabilized read-out circuitry has been built of discrete elements at BUTE (Figure 6). Several measurements have been carried out on nine different SOI-MEMS accelerometers of the same type. Our primary intention was to gain a quantitative characterization of the devices before and after an accelerated lifetime test to allow for comparison. We accomplished a series of measurements on the available devices at a constant excitation. We observed the results of the different devices within each sequence and the results of each device across the series. The former is the indicator of the stability of the sensor process technology, the latter gives information on the stability of the sensors, i. e. their quality of producing an invariant response to a constant excitation.

The charge amplifier (Figure 5.) responsible for the modulation can be described by the following equations:

$$V_y^+ = -V_M \frac{C_S^-}{C_I} = -V_M \frac{C_S - \Delta C}{C_I}$$
$$V_y^- = -V_M \frac{C_S^+}{C_I} = -V_M \frac{C_S + \Delta C}{C_I} \quad (2)$$

Thus it's output voltage is:

$$V_y = V_y^+ - V_y^- = V_M \frac{2\Delta C}{C_I} \quad (3)$$

where $V_M = 1$ V, $C_I \cong 5$ pF, and the value of $\Delta C$ was in the range of 100..1000 fF. The factor of 2 is due to the half bridge arrangement.

The sensing capacitor's ($C_S$) capacitance is 7.048 pF, the change of the capacitance at an acceleration of 1 g is 61.84 fF, which results in a 24 mV change in the output signal's amplitude. This value is amplified by the following stages.

In order to ensure the accuracy of the measurement, the load at the output of the fully differential amplifier needs to be minimalized. Therefore an instrumentation amplifier was inserted before the analog multiplier. This amplifier also performs a differential-to-single-ended conversion. A fourth order low-pass filter is necessary for a precise elimination of the noise at the frequency band of the carrier signal.

The Flicker-noise attenuation of the read-out circuitry proved to be outstanding. The circuit is extremely sensitive to parasitic capacitances however. Throughout the process of the design much respect had to be paid to the reduction of stray capacitances and the prevention of cross-coupling. As to the inevitable parasitic capacitances (e.g. the cables connecting the sensor and the read-out circuitry), all measures had to be taken in order to achieve a symmetric arrangement to allow for differential cancellation. In the course of the measurements the relative shift of the circuit's elements could change the results, hence the set-up had to be fastened to a platform with low dielectric constant. The instruments used for the measurements (power supply, oscilloscope, etc.) had to be connected in a thoughtful way to avoid cross-coupling.

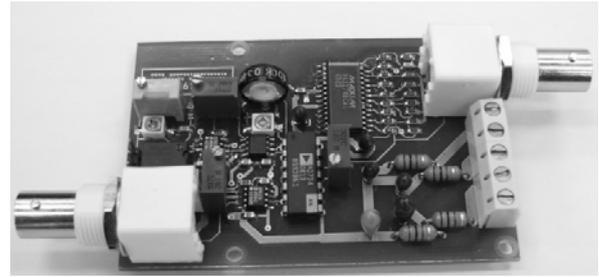

Figure 6 – The chopper stabilized read-out circuitry

Ten series of measurements have been carried out on nine samples. Figure 7 shows the results of the samples in one of the measurement series (No. 10). The average of the output values was 58.44 mV, the standard deviation was 2.79 mV (4.77 %). In order to gather information on the stability of the sensors, the measurement values of the samples across the series of measurements were also observed – this can be seen in Figure 8. The average of the values of sample No.9 was 55.07 mV, the standard deviation was 0.99 mV (1.79 %). The small values of the deviation show that the measurement and the devices were stable.

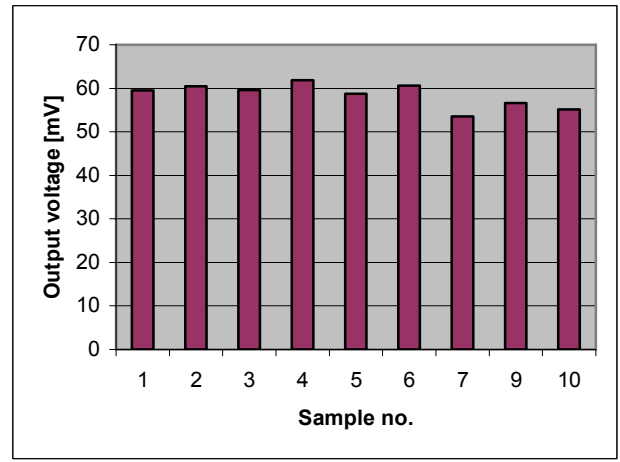

Figure 7 – Result of the samples in a sequence of measurements

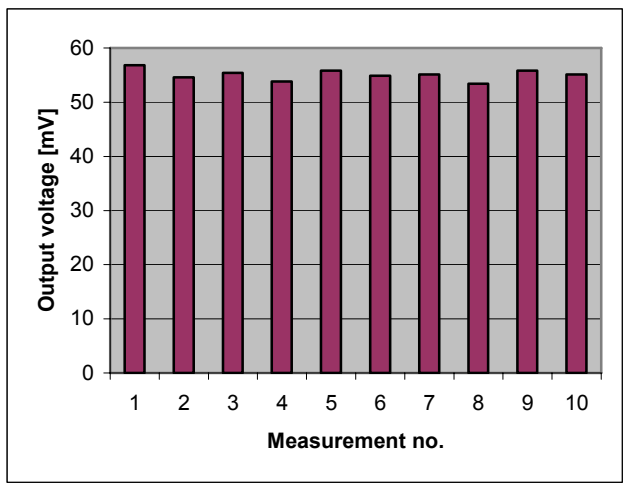

Figure 8 – Values of one sample across a series of measurements

## 3. CONCLUSION

Comparing the two solutions, the following notices have been drawn. The capacitance-to-frequency converter allows measuring a wide range of capacitances with a lower accuracy and stability. This method results in a simple and cheap circuitry in both discrete and integrated forms. In the discrete form, the parasitic capacitances and the temperature-dependency of the elements are the sources of instability. The integrated realization of the circuit is expected to eliminate these defects.

The chopper-stabilized solution provides a more accurate measurement in a substantially smaller capacitance range. This method is applicable for the measurement of small, differential capacitances that may change slowly, and to eliminate the effects of the parasitic capacitances and the Flicker-noise. The chopper-stabilized solution results in a more complex and expensive circuit in discrete and integrated realization.